\documentclass{article}

\usepackage{arxiv}

\usepackage[utf8]{inputenc} 
\usepackage[T1]{fontenc}    
\usepackage{lmodern}        
\usepackage{hyperref}       
\usepackage{url}            
\usepackage{booktabs}       
\usepackage{amsfonts}       
\usepackage{nicefrac}       
\usepackage{microtype}      
\usepackage{lipsum}
\usepackage{graphicx}

\title{AI-Ethics by Design. Evaluating Public Perception on the
Importance of Ethical Design Principles of AI.}

\author{
    Kimon Kieslich
    \href{https://orcid.org/0000-0002-6305-2997}{\includegraphics[scale=0.06]{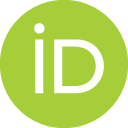}}
   \\
    Department of Social Sciences \\
    Heinrich Heine University \\
  Düsseldorf, Germany \\
  \texttt{\href{mailto:kimon.kieslich@hhu.de}{\nolinkurl{kimon.kieslich@hhu.de}}} \\
   \And
    Birte Keller
    \href{https://orcid.org/0000-0002-3145-5206}{\includegraphics[scale=0.06]{orcid.png}}
   \\
    Department of Social Sciences \\
    Heinrich Heine University \\
  Düsseldorf, Germany \\
  \texttt{\href{mailto:birte.keller@hhu.de}{\nolinkurl{birte.keller@hhu.de}}} \\
   \And
    Christopher Starke
    \href{https://orcid.org/0000-0001-7899-6029}{\includegraphics[scale=0.06]{orcid.png}}
   \\
    Department of Social Sciences \\
    Heinrich Heine University \\
  Düsseldorf, Germany \\
  \texttt{\href{mailto:christopher.starke@hhu.de}{\nolinkurl{christopher.starke@hhu.de}}} \\
  }

\newlength{\csllabelwidth}
\newlength{\cslhangindent}
  {}%
  {\par}
\newenvironment{CSLReferences}[3] 
 {
  \setlength{\parindent}{0pt}
  \ifodd #1 \everypar{\setlength{\hangindent}{\cslhangindent}}\ignorespaces\fi
  \ifnum #2 > 0
  \setlength{\parskip}{#2\baselineskip}
  \fi
 }%
 {}
\usepackage{calc} 

\usepackage{booktabs}
\usepackage{longtable}
\usepackage{array}
\usepackage{multirow}
\usepackage{wrapfig}
\usepackage{float}
\usepackage{colortbl}
\usepackage{pdflscape}
\usepackage{tabu}
\usepackage{threeparttable}
\usepackage{threeparttablex}
\usepackage[normalem]{ulem}
\usepackage{makecell}

\begin{document}
\maketitle

\def\tightlist{}

\begin{abstract}
Despite the immense societal importance of ethically designing
artificial intelligence (AI), little research on the public perceptions
of ethical AI principles exists. This becomes even more striking when
considering that ethical AI development has the aim to be human-centric
and of benefit for the whole society. In this study, we investigate how
ethical principles (explainability, fairness, security, accountability,
accuracy, privacy, machine autonomy) are weighted in comparison to each
other. This is especially important, since simultaneously considering
ethical principles is not only costly, but sometimes even impossible, as
developers must make specific trade-off decisions. In this paper, we
give first answers on the relative importance of ethical principles
given a specific use case --- the use of AI in tax fraud detection. The
results of a large conjoint survey (n=1099) suggest that, by and large,
German respondents found the ethical principles equally important.
However, subsequent cluster analysis shows that different preference
models for ethically designed systems exist among the German population.
These clusters substantially differ not only in the preferred
attributes, but also in the importance level of the attributes
themselves. We further describe how these groups are constituted in
terms of sociodemographics as well as opinions on AI. Societal
implications as well as design challenges are discussed.
\end{abstract}


\hypertarget{introduction}{%
\section{Introduction}\label{introduction}}

Artificial intelligence (AI) has enormous potential to change society.
While the widespread implementation of AI systems can certainly generate
economic profits, policymakers and scientists alike also highlight the
ethical challenges accompanied by AI. Most scholars, politicians, and
developers agree that AI needs to be developed in a human-centric and
trustworthy fashion, resulting in AI that benefits the common good,
respectively the whole society (Berendt, 2019; Cath et al., 2018;
European Commission, 2020; Floridi et al., 2018; Jobin et al., 2019).
Trustworthy and beneficial AI requires that ethical challenges be
considered during all stages of the development and implementation
process. While plenty of work considers ethical AI development, there is
surprisingly little research investigating public perceptions of those
ethical challenges. This lack of citizen involvement is striking because
developing ethical AI aims to be human-centric and of profit for the
whole society. Filling this research gap, we set out to shed light on
public perceptions of ethical principles outlined in ethical guidelines.
Particularly, we investigate how people prioritize different ethical
principles. Accounting for the trade-offs between the different ethical
principles is especially important because maximizing them
simultaneously often proves challenging or even impossible when
designing and implementing AI systems. For instance, aiming for a high
degree of explainability of AI systems can conflict with the ethical
principle of accuracy, since a high degree of accuracy tends to require
complex AI models that cannot be fully understood by humans, especially
laypersons. Thus, taking the goal of ethical AI development seriously
requires decision makers to take the opinions of the (affected) public
into account.

This paper gives first answers to the relative importance of ethical
principles given a specific use case --- here, the use of AI in tax
fraud detection. In a large (\emph{n}=1099) online survey with a
conjoint design, we asked participants to rate different configurations
of tax fraud systems; the proposed systems varied in how they comply
with the seven ethical principles that are most prominent in global AI
guidelines (Jobin et al., 2019). As we aim for high external information
value of our results, we decided not to rely on one specific ethical
guideline, but on the ethical principles that are most referred to on a
global scale.

\hypertarget{ethical-guidelines-of-ai-development}{%
\section{Ethical Guidelines of AI
Development}\label{ethical-guidelines-of-ai-development}}

AI increasingly permeates most areas of people's daily lives, whether in
the form of virtual intelligent assistants such as Alexa or Siri, as a
recommendation algorithm for movie selection on Netflix, or in hiring
processes. Such areas of application are only made possible by the
accumulation of huge amounts of data, so-called Big Data, that people
constantly leave behind in their digital lives. Although these AI-based
technologies aim to take tasks off people's hands and make their lives
easier, collecting and processing personal data is also associated with
major concerns. boyd and Crawford (2012) emphasize the importance of
ethically responsibly handling Big Data. Scandals such as the NSA affair
or Cambridge Analytica have recently caused great public outcry. The
public attention was, therefore, once again drawn more strongly to the
issue of privacy. Policymakers are increasingly reacting to these
concerns. For example, shortly after the Cambridge Analytics scandal
became public, the European Union's General Data Protection Regulation
(GDPR) came into force. This is considered an important step forward in
the field of global political convergence processes and helps to create
a global understanding of how to handle personal data (Bennett, 2018).

However, Big Data not only leads to privacy concerns, but can undermine
transparency for users of online services. This lack of transparency is
further exacerbated by the fact that algorithms are sometimes too
complex for laypersons to understand, which is often referred to as a
black box (Shin and Park, 2019). Questions regarding comprehensibility
and explainability are therefore at the core of algorithmic
decision-making and its outcome (Ananny and Crawford, 2018). These
questions become particularly relevant when algorithms make biased
decisions and systematically discriminate against individual groups of
people. For example, the COMPAS algorithm used by United States (US)
courts systematically disadvantaged black defendants by giving them a
higher risk score for the probability of recidivism than white
defendants (Angwin et al., 2016). In contrast, a hiring algorithm used
by Amazon systematically discriminated against female candidates
(Köchling and Wehner, 2020). Algorithmic discrimination can be caused by
flawed or biased input data or by the mathematical architecture of the
algorithm (Shin and Park, 2019; Zou and Schiebinger, 2018). Thus, AI
systems run in danger of reproducing or even exacerbating existing
social inequalities with detrimental effects for minorities. Such
algorithmic unfairness then leads to the question of who is accountable
for possibly biased decisions by an AI system (Busuioc, 2020;
Diakopoulos, 2016). All of these questions have been extensively
discussed in the fairness, accountability, and transparency in machine
learning (FATML) literature (Shin and Park, 2019). The different
concepts are closely intertwined. For example, Diakopoulos (2016) points
out: ``Transparency can be a mechanism that facilitates accountability''
(p.~58).

To address these concerns and to define common ground for
(self-)regulation, governments, private sector companies, and civil
society organizations have established ethical guidelines for developing
and using AI. The goal is to address the challenges outlined by the
scientific community and thus to ensure so-called ``human-centered AI''
(e.g., Lee et al., 2017; Shneiderman, 2020), or ``human-centric'' AI
(European Commission, 2019). For example, the High-Level Expert Group on
AI (AI HLEG) set up by the European Commission calls for seven
requirements of trustworthy AI: (1) human agency and oversight, (2)
technical robustness and safety, (3) privacy and data governance, (4)
transparency, (5) diversity, non-discrimination and fairness, (6)
societal and environmental well-being, and finally (7) accountability
(European Commission, 2019). Along similar lines, the OECD recommends a
distinction between five ethical principles, namely (1) inclusive
growth, sustainable development, and well-being; (2) human-centered
values and fairness; (3) transparency and explainability; (4)
robustness, security, and safety; and (5) accountability (OECD, 2021).

Some researchers have taken a comparative look at the numerous
guidelines published in recent years and have highlighted which ethical
principles are emphasized across the board (e.g., Hagendorff, 2020;
Jobin et al., 2019). There is widespread agreement on the need for
ethical AI, but not on what it should look like in concrete terms. For
example, Hagendorff (2020) highlights that the requirements for
accountability, privacy, and fairness can be found in 80\% of the 22
guidelines he analyzed. Thus, to a large extent, the guidelines mirror
the primary challenges for human-centric AI discussed in the FATML
literature. At the same time, however, Hagendorff (2020) points out that
it is precisely these principles that can be most easily mathematically
operationalized and thus implemented in the technical development of new
algorithms.

Jobin et al. (2019) conducted a systematic review of a total of 84
ethical guidelines from around the globe, although the majority of the
documents originate from Western democracies. In total, the authors
identify 11 overarching ethical principles, five of which (transparency,
justice and fairness, non-maleficence, responsibility, and privacy) can
be found in more than half of the guidelines analyzed. Also, the
attributes of beneficence and of freedom and autonomy can still be found
in 41 and 34 of the 84 guidelines, respectively. The ethical principles
of trust, sustainability, dignity, and solidarity, on the other hand,
are only mentioned in less than a third of the documents (Jobin et al.,
2019).

In this paper, we focus on the seven most prominent ethical principles,
which are discussed in most of the existing guidelines, analyzed by
Jobin et al. (2019). In addition to the aforementioned principles of
transparency (or explainability), fairness, responsibility
(accountability), and privacy Jobin et al. (2019) list non-maleficence,
freedom and autonomy, and beneficence. They conceive ``general calls for
safety and security'' (p.~394) as non-maleficence. At the core of the
principles lies the requirement for technical security of the system,
for example, in the form of protection against hacker attacks. In this
way, unintended harm from AI should be prevented, in particular
(European Commission, 2019). According to Jobin et al. (2019), the
freedom and autonomy issue addresses, among other things, the risk of
manipulation and monitoring of the process and decisions, as also
addressed by the AI HLEG. In light of this challenge, implementing human
oversight in the decision-making process can ensure that human autonomy
is not undermined and unwanted side effects are thus avoided (European
Commission, 2019). However, decision-making procedures are perceived as
fair when the procedure guarantees a maximum degree of consistency on
the one hand and is free from personal bias on the other (Leventhal,
1980). Therefore, the neutrality of an algorithmic decision --- without
human bias --- might explain why algorithmic decision-making is
perceived as fairer than human decisions (Helberger et al., 2020;
Marcinkowski et al., 2020). Even though these perceptions are context
dependent (Starke et al., 2021), it can be assumed that in some use
cases no human control is desired. For example, this is especially
important to consider when personal bias or, at worst, corruptibility of
human decision makers could be suspected, i.e.~in tax fraud detection
(Köbis et al., 2021). In this sense, the use of AI can lead to less
biased decisions (Miller, 2018). Finally, beneficence refers to the
common good and the benefit to society as a whole. However, reaping this
benefit requires algorithms that do not make any mistakes. The accuracy
of AI is therefore decisive for societal benefit. This is because only a
high level of predictive accuracy or correct decisions made by an AI can
generate maximum benefit (Beil et al., 2019). Accordingly, AI systems
used in medical diagnosis, for instance, can only improve personal and
public health if they operate as accurately as possible (Graham et al.,
2019; Yeasmin, 2019).

While all ethical principles highlighted in the ethical AI guidelines
seem desirable in principle, they can cause considerable challenges in
practice. The reason is that when designing an AI system, it is often
infeasible to maximize the different ethical aspects simultaneously.
Thus, multiple complex trade-off matrices emerge (Binns and Gallo, 2019;
Köbis et al., 2021). Two examples help to illustrate this point. First,
the more available information about a user's wants, needs, and actions,
recommendation algorithms on social media platforms can make helpful and
more accurate recommendations. This information includes private data
about a user, such as the browsing history, but also sensitive data,
such as gender. Collecting this data and simultaneously improving the
recommendation can result in accuracy-privacy (Machanavajjhala et al.,
2011) or accuracy-fairness trade-offs (Binns and Gallo, 2019). Second,
for a company to assess if its hiring algorithm discriminates against
social minorities, it needs to collect sensitive information from their
applicants, such as ethnicity, which may violate fundamental privacy
rights, leading to a fairness-privacy trade-off (Binns and Gallo, 2019).
By adding more variables like transparency, security, autonomy, and
accountability to the mix, highly complex trade-offs between the various
ethical principles emerge.

\hypertarget{public-preferences-for-ai-ethics-guidelines}{%
\section{Public Preferences for AI Ethics
Guidelines}\label{public-preferences-for-ai-ethics-guidelines}}

A human-centric approach requires that AI systems are used ``in the
service of humanity and the common good, with the goal of improving
human welfare and freedom'' (European Commission, 2019: 4). Thus, it is
essential to account for the perceptions of those most affected by
decisions made by algorithmic systems. A recent example from the United
Kingdom (UK) illustrates that violating ethical principles when
designing and implementing AI --- in this case, an automated system that
graded students in schools --- can lead to substantial public outrage
(Kelly, 2021). Empirical research further suggests that perceiving AI as
unethical has detrimental implications for an organization in terms of a
lower reputation (Acikgoz et al., 2020) as well as a higher likelihood
for protests (Marcinkowski et al., 2020) and for pursuing litigation
(Acikgoz et al., 2020). Thus, to address the fundamental question of
which kind of AI we want as a society, detailed knowledge about public
preferences for AI ethics principles is key. A surging strand of
empirical research addresses this question and finds that public
preferences for AI are highly dependent on the context, as well as on
individual characteristics (Pew Research Center, 2018; Starke et al.,
2021). While people perceive algorithms to be acceptable in some domains
(e.g., social media recommendation), they reject them in others (e.g.,
predicting finance scores). Also, judgments about AI hinge considerably
on sociodemographic features, such as age or ethnicity. In the US, a
study by the Pew Research Center (2018) identifies four major concerns
voiced by respondents: (1) privacy violation, (2) unfair outcomes, (3)
removing the human element from crucial decisions, and (4) inability of
AI systems to capture human complexity.

The empirical literature further shows that people largely desire to
incorporate ethical principles advocated for in the legal guidelines.
First, people base their assessment of an AI system on its accuracy. The
seminal study by Dietvorst et al. (2015) finds that people avoid
algorithms after seeing them making a mistake, even if the algorithm
still outperforms human decision makers. Along similar lines, people
lose trust in defective AI systems (Robinette et al., 2017). However,
studies have found that people still follow algorithmic instructions
even after seeing them err (Robinette et al., 2016; Salem et al., 2015).
Second, fairness is a crucial indicator for evaluating AI systems
(Starke et al., 2021). When an AI system is perceived as unfair, it can
lead to detrimental consequences for the institution implementing such a
system (Acikgoz et al., 2020; Marcinkowski et al., 2020). Third,
empirical evidence suggests that keeping humans in the loop of
algorithmic decisions, i. e. ensuring human oversight at least at some
points of the decision-making process, is perceived as fairer
(Nagtegaal, 2021) and more legitimate (Starke and Lünich, 2020) compared
to leaving decisions to algorithms. Fourth, in terms of transparency,
the literature yields mixed results. On the one hand, more openness
about the algorithm is essential to building trust in AI systems
(Neuhaus et al., 2019), involving the users (Kizilcec, 2016), reducing
anxiety, (Jhaver et al., 2018) and increasing user experience (Vitale et
al., 2018). On the other hand, studies show that too much transparency
can impair user experience (Lim and Dey, 2011) and confuse users,
complicating the interaction between humans and AI systems (Eslami et
al., 2018). Fifth, privacy protection can be an essential factor for
evaluating AI systems, leading people to reject algorithmic
recommendations based on personal data (Burbach et al., 2018). However,
other studies suggest that users are often unaware of privacy risk and
rarely use privacy control settings on AI-based devices (Lau et al.,
2018; Zheng et al., 2018). Sixth, empirical research suggests that
people perceive unclear responsibility and liability for algorithmic
decisions as one of the most crucial risks of AI (Kieslich et al.,
2020). Furthermore, accountability and clear regulations are viewed as
highly effective countermeasures to algorithmic discrimination (Kieslich
et al., 2020). Along similar lines, other studies found that perceptions
of accountability increase people's satisfaction with algorithms (Shin
and Park, 2019) as well as their trust (Shin et al., 2020). Lastly, in
terms of security, people consider a loss of control over algorithms a
crucial risk of AI systems (Kieslich et al., 2020).

Only a few studies, however, compare the influence of different ethical
indicators on people's preferences. In several studies, Shin and
colleagues tested the effects of three crucial aspects of ethical AI:
fairness, transparency, and accountability. The results, however, are
mixed. While fairness had the most substantial impact on people's
satisfaction with algorithms (followed by transparency and
accountability) (Shin and Park, 2019), transparency was the strongest
predictor for people's trust in algorithms (followed by fairness and
accountability) (Shin et al., 2020). Another study found that
explainability has the most decisive influence on algorithmic trust
(Shin, 2020). However, to the best of our knowledge, no empirical study
has looked at different trade-off matrices between the various ethical
principles and investigated people's preferences for single principles
at the expense of others. Therefore, we propose the following research
question:

RQ1: How do varying degrees of consideration of ethical principles in
the design of an AI-based system influence the public's preference for
prioritization among them?

However, considering the diversity of social settings and beliefs among
society, it is probable that there are trade-off differences among the
public concerning the prioritization of ethical principles, respectively
ethical preference patterns of AI systems. Hence, we ask the following
research question:

RQ2: Which preference patterns of ethical principles are present in the
German public?

The literature suggests that human-related factors influence the
perception of AI systems. For example, empirical studies have found that
age (Grgić-Hlača et al., 2020; Helberger et al., 2020; Vallejos et al.,
2017), educational level (Helberger et al., 2020), self-interest
(Grgić-Hlača et al., 2020; Wang et al., 2020), familiarity with
algorithms (Saha et al., 2020), and concerns about data collection
(Araujo et al., 2020) have effects on the perception of algorithmic
fairness. Hancock et al. (2011) performed a meta-analysis of factors
influencing trust in human-robot interaction and identified, among
others, demographics and attitudes toward robots as possible predictors.
Subsequently, we elaborate on this literature and test for differences
among human-related factors concerning the emerging ethical design
patterns. Hence, we ask RQ3.

RQ3: Which characteristics do people who favor a specific ethical design
of AI systems share?

\hypertarget{method}{%
\section{Method}\label{method}}

\hypertarget{sample}{%
\subsection{Sample}\label{sample}}

The data were collected via the online access panel (OAP) of the market
research institute \emph{forsa} between March 16 and March 25, 2021. The
OAP is representative of the German population above 18 years of age,
which at least occasionally uses the Internet. Respondents from the
panel were randomly invited to participate in the survey, with each
panelist having the same chance to be part of the sample. Altogether,
1,204 people participated in the survey.

We cleaned the data according to three criteria: (1) low response time
for the entire questionnaire (1 SD under average time), (2) high number
of missing data in the entire questionnaire (2 SD above average number
of missing data), and (3) low reading time of the introduction text for
the conjoint analysis (under 20 seconds reading time identified through
a pre-test). Participants were excluded when all criteria were met.
Consequently, one participant was excluded. Additionally, we excluded
all respondents who rated all proposed systems in the conjoint analysis
equally (\emph{n}=104). This data cleaning step was crucial, since those
respondents showed no preferences for any configuration and,
methodologically speaking, for those respondents, no variance can be
explained in the conjoint analysis.

After data cleaning, 1099 cases remained. In total, our sample consisted
of 593 (54\%) women and 503 (45.8\%) men, while 3 (0.3\%) indicated
binary.. The average age of the respondents was 47.1 (\emph{SD}=16.7).
Furthermore, regarding education level, 192 (17.5\%) hold a low, 362
(32.9\%) hold a middle and 540 (49.1\%) hold a high educational
degree.\footnote{Five persons (0.05\%) in the sample didn't indicate
  their educational level.}

\hypertarget{procedure}{%
\subsection{Procedure}\label{procedure}}

Initially, respondents were asked to answer several questions concerning
their perception and opinion on AI. To evaluate the preference of
ethical principles in the design of AI systems, we integrated a conjoint
survey with seven attributes in the survey. The use case addresses an
AI-based tax fraud detection system. Such systems are already in use in
many European countries, e.g., France, the Netherlands, Poland, and
Slovenia (Algorithm Watch, 2020) and also in the state of Hesse in
Germany (Institut für den öffentlichen Sektor, 2019). In our study,
respondents were presented with a short text (179 words) describing the
use case of AI in tax fraud detection. The text stated that these
systems can be designed differently. Then, we described the seven most
prominent principles in ethical AI guidelines that we derived from the
review paper by Jobin et al. (2019): explainability (as measurement for
the dimension ``transparency''), fairness, security (as measurement for
the dimension ``non-maleficence''), accountability (as measurement for
the dimension ``responsibility''), accuracy (as measurement for the
dimension ``societal well-being''), privacy, and limited machine
autonomy (for exact wording of the attributes, see Table 1). Notably, we
chose to include machine autonomy as we assumed that in the special case
of tax fraud detection, no human oversight might be preferred due to
possible bias reduction. In the following, the ethical principles are
also called ``attributes''.\footnote{As the ethical principles outlined
  by Jobin et al. (2019) are rather broad, we consulted the guidelines
  of the EU commission (European Commission, 2019) for some formulations
  of the attributes. We take this measure as the German AI strategy is
  oriented on the EU guidelines and we aimed for a comprehensible design
  of the attributes.}

After reading the short introductory text, respondents were told that an
AI system can have different configurations of the different ethical
principles. If the system satisfied a principle, it was indicated with a
green tick; if the property was not met, it was marked with a red cross.
Respondents were presented with a total of eight cards showing different
compositions of AI systems in randomized order. The configurations only
varied whether the seven ethical principles were satisfied (see Table 4
in the appendix). For each card, we asked respondents to indicate how
much they preferred the configuration of ethical principles shown on the
card. At the end of the questionnaire, respondents had to indicate some
sociodemographic information.

\begin{table}

\caption{\label{tab:unnamed-chunk-2}Desciption of the attributes}
\centering
\begin{tabular}[t]{l>{\raggedright\arraybackslash}p{35em}}
\toprule
Ethical Principle & Description\\
\midrule
Explainabilty & \emph{Explanation of the decision}: each/any person concerned is explained in a generally understandable way why the system has classified him/her as a potential tax fraudster.\\
Fairness & \emph{No systematic discrimination}: No persons (groups) are systematically disadvantaged by the automated tax investigation.\\
Security & \emph{State-of-the-art security technology}: The protection of the computer system against hacker attacks is always kept up to date with the latest security technology.\\
Accountability & \emph{Full responsibility with the tax authority}: Should the automated tax investigation system lead to false accusations, the responsible tax authority bears full responsibility for any damage incurred.\\
Accuracy & \emph{Virtually no errors in decision-making}: The automated identification of tax fraud by the computer system works almost without errors.\\
\addlinespace
Privacy & \emph{Exclusively earmarked use of data}: Only the necessary data is used by the automated tax investigation system. Any other use of the considered data is excluded.\\
Machine Autonomy & \emph{No human supervision}: The identification of suspicious cases remains the sole responsibility of the automated tax investigation system.\\
\bottomrule
\end{tabular}
\end{table}

\hypertarget{measurement}{%
\subsection{Measurement}\label{measurement}}

\hypertarget{conjoint-design}{%
\subsubsection{Conjoint-Design}\label{conjoint-design}}

The strength of conjoint analysis lies in its ability to analyze a
variety of possible attributes simultaneously (Green et al., 2001). This
is particularly relevant for attributes that can potentially offset each
other in reality, as argued in the trade-offs of the ethical principles.
While asking for the approval of the principles separately is likely to
yield high scores across the board, conjoint analysis forces respondents
to make a choice between the imperfect configurations of the principles.
Furthermore, conjoint surveys can also be conducted with a partial
factorial design. Thus, it is possible to predict respondents'
preferences for all combination possibilities, even if they only rate a
small fraction of them. As described above, we treated the seven most
prominent ethical design principles outlined by Jobin et al. (2019) as
attributes (transparency, fairness, non-maleficence, responsibility,
beneficence, privacy, autonomy). We chose sub-codes for some ethical
principles to tailor the broad concepts to our use case of tax fraud
detection. As attributed levels, we simply marked if an ethical
principle was complied with or not.

To determine the different compositions of the cards used in our study,
we calculated a fractional factorial design using a standard ``order''
allocation method and random seed. This method produces an orthoplan
solution in which combinations of attributes are well balanced.

\hypertarget{measures}{%
\subsubsection{Measures}\label{measures}}

\emph{System approval}. The approval of each system configuration was
measured using a single item on a seven-point Likert scale (1=``do not
like the presented system at all''; 7=``really like the presented
system'').

\emph{Interest in AI}. To gauge people's interest in AI, respondents
were asked to rate four items on a five-point Likert scale (1=``not true
at all''; 5=``very true''); for instance, ``In general, I am very
interested in artificial intelligence'' (see Table 5 in appendix for
exact wording). We used the four items to compute a highly reliable mean
index (\emph{M}=2.79; \emph{SD}=1.07; \(\alpha\)=0.94). Scale and
wording was adopted from the \emph{Opinion Monitor Artificial
Intelligence} (Meinungsmonitor Künstliche Intelligenz, 2021).

\emph{Acceptance of AI in domains}. Respondents were asked whether they
support the use of AI in 14 different domains on a five-point Likert
scale (1=``no support at all''; 5=''totally support``). For every
domain, we grouped the support values (4 and 5) as support for AI in the
specific domain. Afterwards, we calculated an acceptance sum index;
thus, the sum index ranges from 0=``support in none application domain''
to 14=``support in all application domains'', \emph{M}=3.96
(\emph{SD}=2.96). The measurement was adapted from the \emph{Opinion
Monitor Artificial Intelligence} (Meinungsmonitor Künstliche
Intelligenz, 2021).

\emph{Risk awareness of AI}. We measured risk awareness of AI with three
items by asking respondents: ``You can associate both advantages and
disadvantages with artificial intelligence. Completely independent of
how big you think a possible benefit is: How great do you think the risk
posed by artificial intelligence is?'' Respondents gave their opinion
toward their risk perception for the whole society and themselves, as
well as for family and friends. The items were measured on a ten-point
Likert scale (1=``no risk at all'' to 10=``very high risk''). We adapted
the measurement by Liu and Priest (2009) and calculated a highly
reliable mean index (\emph{M}=4.9; \emph{SD}=1.92; \(\alpha\)=0.91).

\emph{Opportunity awareness of AI}. Along similar lines, respondents
were asked to rate three items on a ten-point Likert scale to the
question: ``Completely independent of the risk, how great do you think
is the benefit to be gained from artificial intelligence.'' Again, they
had to rate the benefit perception for the whole society, themselves,
and friends and family. We adapted the measurement by Liu and Priest
(2009) and computed a highly reliable mean index (\emph{M}=5.82;
\emph{SD}=1.76; \(\alpha\)=0.87).

\emph{Trust in AI}. We measured trust in AI with four items to the
question ``How much do you trust systems of artificial intelligence
already today\ldots{}'' on a ten-point Likert scale (1=''do not trust at
all'' to 10=``trust completely''). An example item read as follows:
``\ldots recognize patterns in large data sets''. We calculated a
reliable mean index (\emph{M}=5.81; \emph{SD}=1.73; \(\alpha\)=0.76).
The question wording was adapted from Lee (2018). The items are based on
the dimensions proposed by Kieslich et al. (2021).

\hypertarget{results}{%
\section{Results}\label{results}}

All calculations were performed in R (V4.0.3). The analysis code,
including R-packages used, is available on request.

\hypertarget{relative-importance-of-ethical-principles}{%
\subsection{Relative Importance of Ethical
Principles}\label{relative-importance-of-ethical-principles}}

To answer RQ1, we calculated a conjoint analysis in R. In particular, we
computed linear regressions for every respondent with the attributes as
independent variables (dummy coded) and the ratings of the cards as the
dependent variable. Thus, 1,099 regression models were calculated to
show the preferences of every respondent; the regression coefficients
are called the part-worth values (Backhaus et al., 2016; Härdle and
Simar, 2015). Then, we computed the average score of the regression
coefficients to retrieve the preferences of ethical design of the German
population.

\hypertarget{part-worth-of-attributes}{%
\subsection{Part-Worth of Attributes}\label{part-worth-of-attributes}}

Predictably, all regression coefficients (part-worths) were positive
(\(\sf{b_{Accountability}}\)=0.8; \(\sf{b_{Accuracy}}\)=0.64;
\(\sf{b_{Explainability}}\)=0.57; \(\sf{b_{Fairness}}\)=0.66;
\(\sf{b_{Autonomy}}\)=0.32; \(\sf{b_{Privacy}}\)=0.66;
\(\sf{b_{Security}}\)=0.66). Hence, the compliance with every ethical
principle, except for limited machine autonomy, positively influences
the approval rating of an AI system. As mentioned earlier, machine
autonomy can be seen as conducive to objectivity in some cases. Hence,
it can be preferred to human oversight. In the given case, the
respondents aim at average for a solution where tax fraud is arguably
detected unbiasedly.

We further zoomed in on the differences between the importance of
satisfying the ethical principles, or, in other words, people's
preferences for some ethical principles over others. For that, we
calculated the importance weights for each attribute (see Figure 1).
Importance weights can be obtained by dividing each mean attribute
part-worth by the total sum of the mean part-worths.

The results suggest that accountability is, on average, perceived as the
most important ethical principle. Fairness, security, privacy, and
accuracy are on average equally important to the respondents.
Explainability of AI-systems is slightly less important. Lastly, machine
autonomy is least important for the respondents. Thus, the importance
weights of the attributes are, overall, relatively balanced in the
aggregate.

\begin{figure}
\centering
\includegraphics{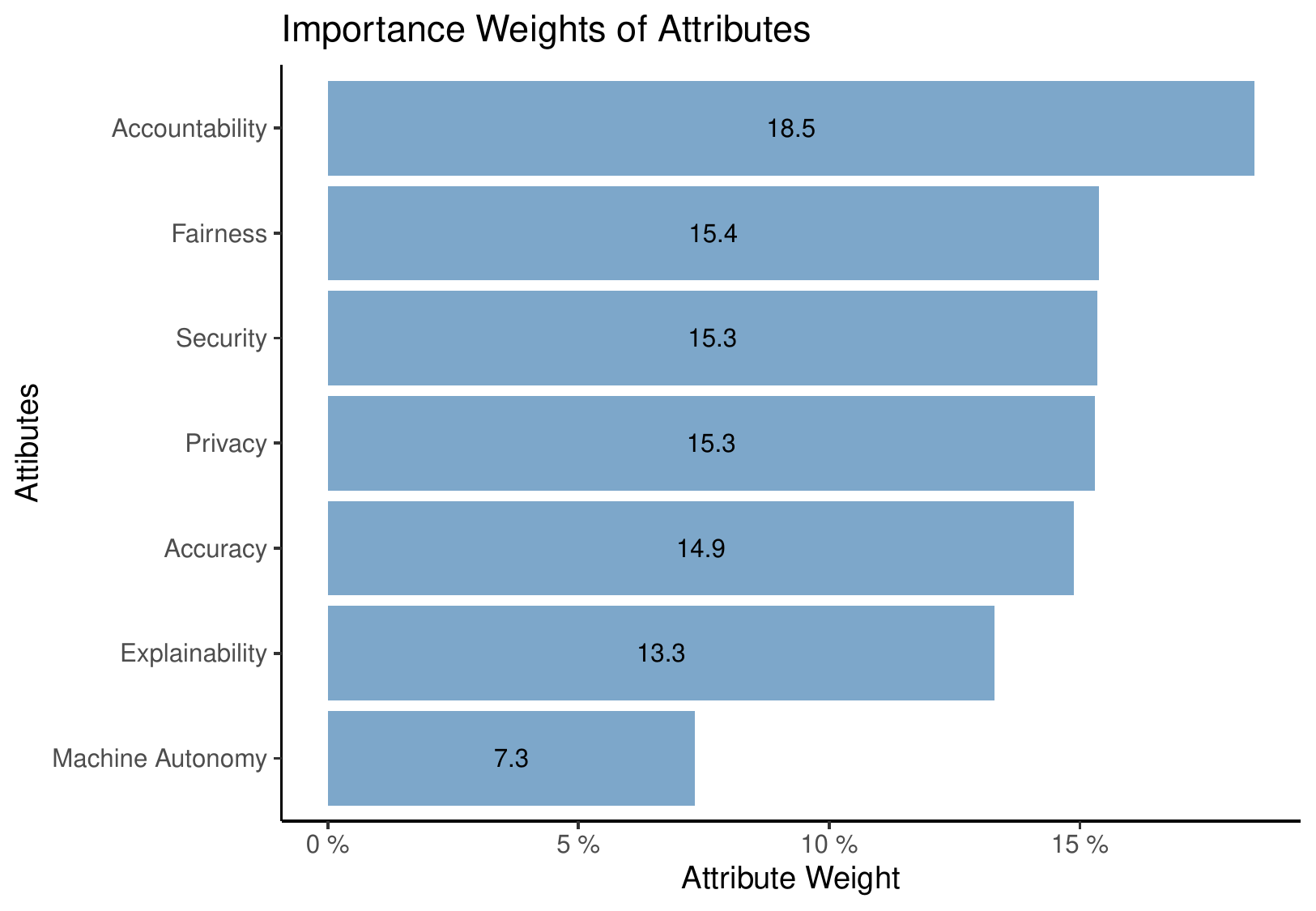}
\caption{Overall Importance Weights of Attributes}
\end{figure}

\hypertarget{preference-patterns-among-the-public}{%
\subsection{Preference Patterns Among the
Public}\label{preference-patterns-among-the-public}}

To answer RQ2 and RQ3, we conducted k-means clustering in R. K-means
clustering is a method used to split observations into k mutually
exclusive groups, called clusters, whereby group members within a group
are as similar as possible and as dissimilar as possible from other
groups (Boehmke and Greenwell, 2020). Thus, k-means clustering provides
solutions for a differentiation of respondents based on a given set of
properties.

We used respondents' regression coefficients as cluster-forming
variables. The number of clusters was determined using the
within-cluster sum of square (``elbow'') method with Euclidean distance
measure. Euclidean distance measure was chosen since the cluster
variables follow a Gaussian distribution and have few outliers. The
results suggest a solution of k=5 or k=11 clusters. Since we aim for a
comprehensible cluster solution and k is commonly determined on
convenience (Boehmke and Greenwell, 2020), we decided to choose the
five-cluster solution in our analysis, as it is clearer to interpret and
allows for further description and comparisons of the groups.
Afterwards, we computed the k-mean clusters using the algorithm of
Hartigan and Wong (1979) using 20 different starting points.

Figure 2 shows the preference profiles of the five cluster groups. The
yellow group includes people who do not seem to care much about the
ethical design of systems. The purple group values fairness, accuracy,
and accountability. The green group demands privacy, security, and
accountability. The blue group considers all ethical principles equally
important. Finally, the main characteristic of people belonging to the
red group is described through high disapproval of machine autonomy.

\begin{figure}
\centering
\includegraphics{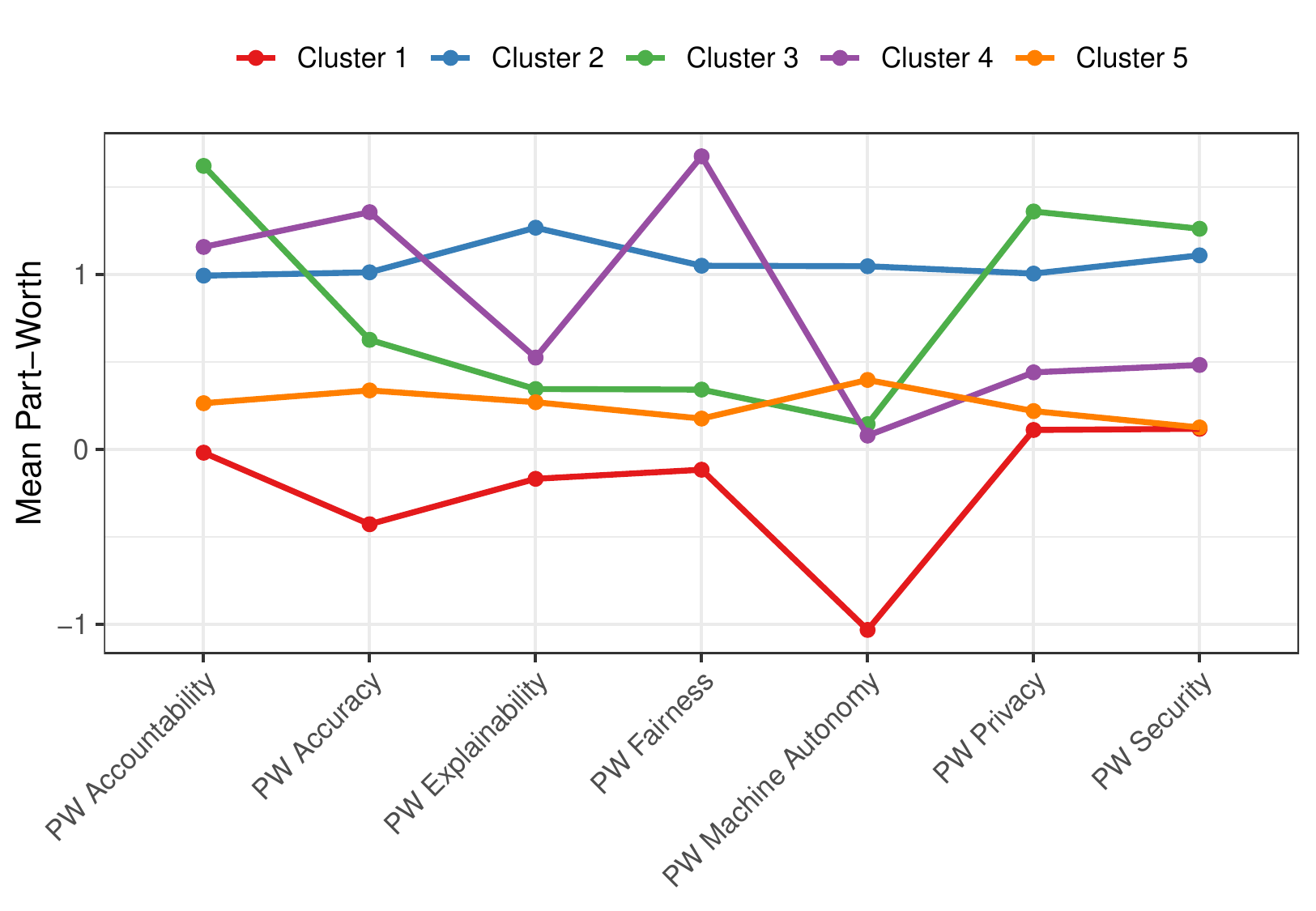}
\caption{Preference Profiles}
\end{figure}

In the next step, we labeled the clusters and calculated the cluster
sizes. Cluster 1 (red) was labeled as ``Human in the Loop'' cluster 2
(blue) as ``Ethically Concerned'' cluster 3 (green) as ``Safety
Concerned,'' cluster 4 (purple) as ``Fairness Concerned'', and cluster 5
(yellow) as ``Indifferent''. Table 2 depicts the average approval
ratings for each cluster group per card and in total across all cards.

The results show that the largest group cluster consists of people who
treat all ethical principles equally and highly important, \emph{n}=345
(31.39\%). Hence, people in the \emph{Ethically Concerned} group
appreciate systems that satisfy all ethical principles. Otherwise, the
approval ratings are quite low.

In contrast, the second largest group consists of people whose system
approval ratings are only slightly affected by an ethical design of an
AI-system, \emph{n}=267 (24.29\%). We label them as \emph{Indifferent}.
Those people do not seem to care much about the ethical design of the
system. However, persons in this cluster group show a medium acceptance
for all presented systems.

A total of 167 (15.2\%) respondents were considered as \emph{Safety
Concerned}. For those, AI systems must be safe, privacy has to be
protected, and the responsibility of a specific entity has to be
ensured. These ethical principles are far more important than fairness,
accuracy, or explainability. Across all presented systems, the approval
ratings are on a low to medium level.

The group of \emph{Fairness Concerned} consists of 166 (15.1\%)
respondents who especially consider fairness and accuracy to be the
important principles, whereas privacy and security hardly affected a
positive rating. The \emph{Fairness Concerned} are quite skeptical
toward the presented systems if they do not follow their demanded
ethical principles.

In the fifth cluster, people oppose machine autonomy and accordingly
strive for human control, \emph{n}=154 (14.01\%). We term this group of
respondents \emph{Human in the Loop} as limited machine autonomy is the
only factor that highly affects the ratings of the AI systems. Hence,
for this group, it is relevant to build systems that are under human
control. However, approval of the presented systems is on average on a
medium level.

\begin{table}

\caption{\label{tab:unnamed-chunk-5}Mean Card Ratings per Group}
\centering
\resizebox{\linewidth}{!}{
\begin{tabular}[t]{lrrrrrrrrr}
\toprule
Group Name & Card A & Card B & Card C & Card D & Card E & Card F & Card G & Card H & Average Rating\\
\midrule
Human in the Loop & 4.19 & 3.14 & 3.01 & 2.82 & 3.32 & 3.87 & 4.56 & 3.79 & 3.59\\
Ethically Concerned & 2.28 & 2.17 & 2.03 & 6.34 & 1.96 & 2.14 & 1.97 & 1.91 & 2.60\\
Safety Concerned & 2.38 & 2.54 & 2.47 & 6.14 & 2.28 & 2.77 & 4.68 & 3.02 & 3.29\\
Fairness Concerned & 3.02 & 2.10 & 2.25 & 6.05 & 2.53 & 2.66 & 2.42 & 4.52 & 3.19\\
Indifferent & 3.18 & 3.54 & 3.46 & 4.40 & 3.40 & 3.43 & 3.21 & 3.38 & 3.50\\
\bottomrule
\end{tabular}}
\end{table}

\hypertarget{cluster-description}{%
\subsection{Cluster Description}\label{cluster-description}}

We address RQ3 by describing the five cluster groups based on several
characteristics, which we group into two categories: socio-demography
(\emph{age}, \emph{educational level}) and AI opinions (\emph{interest},
\emph{acceptance of AI in momains}, \emph{risk awareness of AI},
\emph{opportunity awareness of AI}, \emph{trust in AI}). We only
included those respondents (\emph{n}=913), who answered all included
variables (no missing values). In the first step, we calculated the mean
values for each explanatory variable for each cluster group. To test for
significant mean differences between the clusters, we ran a MANOVA with
the cluster group as the independent grouping variable and the seven
characteristics outlined above as dependent variables. We checked the
assumptions and found homogeneity of variance-covariance matrices using
Box's M test, \emph{M}=137.62, \emph{p}=0.05. As Box's M test is very
sensitive, values lower than .001 are considered to be not trusted
(Field, 2011). Further, we tested for normal distribution of the
dependent variables with visual inspection and multivariate Shapiro-Wilk
test. The Shapiro-Wilk test showed a significant deviance from
normality, \emph{W}(913)=0.98, \emph{p}=0. Moreover, visual inspection
revealed that the data were non-normal distributed. However, MANOVA is
rather robust to a violation of normal distribution (Field, 2011). We
used Pillais' Trace test statistic, as it is the most robust test for
violations of the underlying assumptions (Field, 2011).

As the MANOVA shows statistical significance, \emph{V}=0.11, \emph{F}(4,
908)=3.74, \emph{p}=0, we performed subsequent ANOVA analyses for each
of the dependent variables. Further, post-hoc tests with Tukey-HSD
correction were used to test for mean differences between the groups for
every dependent variable (see Table 3).

\begin{table}

\caption{\label{tab:unnamed-chunk-6}Cluster Descritpion}
\centering
\resizebox{\linewidth}{!}{
\begin{tabular}[t]{llllllrr}
\toprule
  & Human in the Loop & Ethically Concerned & Safety Concerned & Fairness Concerned & Indifferent & F & p\\
\midrule
\addlinespace[0.3em]
\multicolumn{8}{l}{\textbf{Sociodemographics}}\\
\hspace{1em}Age & 49.52 (15.63)\textsuperscript{a} & 44.87 (15.87)\textsuperscript{ab} & 46.28 (16.48)\textsuperscript{a} & 40.54 (17.66)\textsuperscript{b} & 49.46 (16.66)\textsuperscript{a} & 8.18 & 0.00\\
\hspace{1em}Educational Level & 2.24 (0.76)\textsuperscript{c} & 2.49 (0.68)\textsuperscript{ab} & 2.34 (0.76)\textsuperscript{bc} & 2.60 (0.65)\textsuperscript{a} & 2.13 (0.74)\textsuperscript{c} & 12.72 & 0.00\\
\addlinespace[0.3em]
\multicolumn{8}{l}{\textbf{AI Opinions}}\\
\hspace{1em}Interest & 2.77 (1.11)\textsuperscript{ab} & 3.00 (1.10)\textsuperscript{a} & 2.97 (0.95)\textsuperscript{ab} & 3.01 (0.95)\textsuperscript{a} & 2.71 (1.00)\textsuperscript{ab} & 3.44 & 0.01\\
\hspace{1em}Acceptance & 3.58 (2.79)\textsuperscript{c} & 4.52 (3.03)\textsuperscript{ab} & 4.22 (2.77)\textsuperscript{bc} & 5.15 (2.79)\textsuperscript{a} & 3.71 (2.78)\textsuperscript{c} & 7.83 & 0.00\\
\hspace{1em}Risk Awareness & 5.28 (1.93)\textsuperscript{ab} & 4.82 (1.94)\textsuperscript{ab} & 4.71 (2.04)\textsuperscript{bc} & 4.18 (1.68)\textsuperscript{c} & 5.29 (1.78)\textsuperscript{a} & 9.27 & 0.00\\
\hspace{1em}Opportunity Awareness & 5.55 (1.69)\textsuperscript{bc} & 6.04 (1.79)\textsuperscript{ab} & 5.98 (1.77)\textsuperscript{abc} & 6.18 (1.65)\textsuperscript{a} & 5.54 (1.80)\textsuperscript{c} & 4.75 & 0.00\\
\hspace{1em}Trust in AI & 5.50 (1.75)\textsuperscript{b} & 6.05 (1.60)\textsuperscript{a} & 5.91 (1.58)\textsuperscript{ab} & 6.38 (1.48)\textsuperscript{a} & 5.57 (1.95)\textsuperscript{b} & 7.39 & 0.00\\
\bottomrule
\multicolumn{8}{l}{\textsuperscript{*} MANOVA significant using Pillais' Trace Test Statistic, p<.05.}\\
\multicolumn{8}{l}{\textsuperscript{\dag} Cells show mean values and standard deviation (in brackets) for the cluster groups.}\\
\multicolumn{8}{l}{\textsuperscript{\ddag} F and p show the significane of the subsquent ANOVAs performed.}\\
\multicolumn{8}{l}{\textsuperscript{\S} Means in a row without a common superscript (a-c) letter differ (p< 0.05) as analyzed by the ANOVA and the TUKEY post-hoc test.}\\
\end{tabular}}
\end{table}

The ANOVA results show that the clusters significantly deviate from each
other on all analyzed characteristics. In the following, we will
describe the profile of each cluster group in further detail. All mean
scores of all variables are displayed in Table 3 to visualize group
comparisons.

\hypertarget{human-in-the-loop}{%
\subsubsection{Human in the Loop}\label{human-in-the-loop}}

The \emph{Human in the Loop} group overwhelmingly demands human control
and, thus, is strongly opposed to machine autonomy. Persons belonging to
this group tend to be older and less educated. Regarding AI opinions,
they are rather uninterested in AI and have a low acceptance of AI
technologies. Moreover, they are comparatively more aware of risks, have
quite low benefit perceptions, and low levels of trust in AI.

\hypertarget{ethically-concerned}{%
\subsubsection{Ethically Concerned}\label{ethically-concerned}}

People who demand high standards on all ethical principles are
comparatively young and well educated. They also have a high interest
and trust in AI systems. Furthermore, they tend to accept AI and have a
relatively high benefit perception and a relatively low risk perception.

\hypertarget{safety-concerned}{%
\subsubsection{Safety Concerned}\label{safety-concerned}}

Respondents belonging to the \emph{Safety Concerned} group are located
in between the other groups regarding the sociodemographic variables and
AI opinions. They are, of average age and education. Furthermore, they
are somewhat interested in AI, accept AI in some application domains,
are medium risk and benefit aware, and trust AI systems to an average
extent.

\hypertarget{fairness-concerned}{%
\subsubsection{Fairness Concerned}\label{fairness-concerned}}

The \emph{Fairness Concerned} group, which is concerned with the
accuracy and fairness of AI systems, is comparatively young and well
educated. Out of all clusters, the \emph{Fairness Concerned} also
perceive the lowest risks and the greatest benefits of AI. They further
have the highest trust in AI systems, are most accepting of AI, and are
one of the groups with the highest interest in AI.

\hypertarget{indifferent}{%
\subsubsection{Indifferent}\label{indifferent}}

The \emph{Indifferent} can be described --- together with the
\emph{Human in the Loop} --- as the group with the most negative
opinions on AI. People who do not demand ethically designed systems have
relatively low acceptance, low benefit perceptions, and little trust in
AI. Further, they have a high risk awareness and are comparatively
uninterested in AI.

\hypertarget{discussion}{%
\section{Discussion}\label{discussion}}

This study sheds light on an under-researched area of ethical AI, namely
the public perceptions of ethical challenges that come along with
developing algorithms. On one hand, existing research focuses on
normative and legal considerations for how ethical AI systems should be
designed; on the other hand, the computer science literature elaborates
on how ethical AI systems can be designed. However, empirical studies
focusing on people's assessment of ethical principles are relatively
rare. However, as argued in this paper, accounting for the perceptions
of those affected by AI systems is vital for a human-centric approach to
AI.

Thus, we investigated opinions about the ethical design of AI systems by
jointly considering different essential ethical principles and shedding
light on their relative importance (RQ1). We further explored different
preference patterns (RQ2) and how these patterns can be explained in
terms of sociodemographics as well as AI-related opinions (RQ3).

\hypertarget{from-ethical-guidelines-to-legal-frameworks}{%
\subsection{From Ethical Guidelines to Legal
Frameworks?}\label{from-ethical-guidelines-to-legal-frameworks}}

Our results show that regarding the relative importance of ethical
principles no big differences among the German public exist. However, we
find a slight accentuation of \emph{accountability} as the most
important ethical principle; moreover, the respondents consider
\emph{limited machine autonomy} slightly less important than the other
ethical principles. Initially, these aggregate results indicate a
balanced view on ethical AI. None of the ethical principles are strongly
preferred over the other, leading to the conclusion that German citizens
seem to have no critical blind spots. For a good rating of an AI system,
all ethical principles are more or less equally important. Hence,
developers and organizations should not neglect some ethical principles,
while emphasizing others. Based on these results, it seems that
compliance with multiple ethical principles is important for an AI
system to receive a positive rating.

Thus, ethical guidelines are not only present in a vacuum, but also
address the needs of the public. In the case of German citizens,
accountability is foremost demanded. In the context of our study,
accountability is equal to liability; hence, there is a need for a clear
presentation of an actor, who can be accounted for losses and who --- in
the end --- can be regulated. This is in line with empirical evidence
showing that legal regulations are perceived not only as effective, but
also as demanded countermeasures against discriminatory AI (Kieslich et
al., 2020). As AI technology is considered a potential risk or even
threat --- at least among a share of the public (Kieslich et al., 2021;
Liang and Lee, 2017) --- setting up a clear legal framework for
regulation might be a way to further enhance trust and acceptance toward
AI. In this respect, the European Union (EU) has already taken on a
pioneering role, as the EU commission recently proposed a legal
framework for the handling of AI technology (European Commission, 2021).
With this, they set up a classification framework for high-risk
technology and even list specific applications that should be closely
controlled or even banned. Considering the results of our study, this
might be a fruitful way to include citizens, specifically, if it is made
clear who takes responsibility for poor decisions made by AI systems.
Besides, it is the articulated will of the European Commission to put
humans at the center of AI development. Our empirical results suggest
that ethical design matters and --- if the EU takes their goals
seriously --- ethical challenges should play a major role in the future.
Strictly speaking, ethical AI thus primarily requires regulatory
political or legal actions. Hence, the implementation of ethical AI is a
political task, which must not necessarily include computer scientists.
However, from our results, we can also draw conclusions for the ethical
design of AI systems in a technological sense.

\hypertarget{ethical-design-and-demands-of-potential-stakeholder-groups}{%
\subsection{Ethical Design and Demands of Potential Stakeholder
Groups}\label{ethical-design-and-demands-of-potential-stakeholder-groups}}

Our results also suggest that citizens value ethical principles
differently. After clustering the respondents' preferences, we found
five different groups that differ considerably in their preferences for
ethical principles. This suggests that there might not be a universal
understanding and balance of the importance of ethical principles in the
German population. People have different demands and expectations
regarding the ethical design of AI systems. Thus, these different
preference patterns have implications for the (technical) design and
implementation of AI systems. For example, the \emph{Fairness Concerned}
group should be addressed in different ways than the \emph{Safety
Concerned} or the \emph{Human in the Loop} groups. Several studies have
already been conducted on the inclusion of stakeholders in the design
process, especially for fairness (Vallejos et al., 2017; Webb et al.,
2018).

Concerning the results of our study, for example, given the case of an
algorithmic admission system in universities (Dietvorst et al., 2015),
system requirements articulated by the affected public (in this case,
students) might widely differ from those of a job seeker categorization
system (e.g., the algorithmic categorization system used by the Austrian
jobservice (AMS)) (Allhutter et al., 2020). While students supposedly
are younger, well educated, and more interested in AI, those affected by
a job seeker categorization system are supposedly older and feel less
positive about AI. Our results suggest that operators of AI systems
should address the needs of the stakeholders differently if aiming for
greater acceptance. For the admission system, it might be useful to
highlight that such systems are precise and treat students equally,
since students --- based on their group characteristics --- primarily
belong to the group of the \emph{Fairness Concerned}. For the job seeker
categorization system, on the other hand, it might be more promising to
focus on safety issues or the presence of human responsibility, as they
may be assigned to the group of \emph{Safety Concerned} or the
\emph{Human in the Loop}. It should be noted that we explicitly
highlight that we believe that every ethical design principle is of
great importance and that developers should address all issues
accordingly. We simply outline that communication about such systems
could differ concerning the affected public.

Notably, there is also a group of people of substantial size (the
\emph{Indifferent}), who are only slightly concerned with the ethical
design of AI systems. This group does not oppose AI systems in general
(in fact, they have on average the highest approval scores of all
cluster groups for the presented systems), but they are not affected by
compliance with ethical principles. This might be somewhat problematic,
since this group arguably will not set high expectations for companies
that develop AI systems. For example, Elzayn and Fish (2020) showed that
achieving \emph{fairness} in AI systems is very costly and that the
market does not reward putting a massive amount of money in collecting
data of marginalized groups, whether for monopolists or under
competition. This becomes more alarming when considering the share of
the \emph{Indifferent} in society (24\%). One might assume that the
combination of lack of reward for ethical principles by the market and a
potential lack of public outcry --- at least in some parts of society
--- might lead to a sloppy implementation of ethical principles in
practice. This is especially important to consider because ethical
considerations are often left out of software development (McNamara et
al., 2018). Again, Elzayn and Fish (2020) propose policy solutions to
tackle this issue. Besides policy actions, organizations that are
concerned with the ethical design of AI (e.g., \emph{Algorithm Watch})
could actively reach out to the \emph{Indifferent} and try to create
awareness of the consequences of non-compliance with ethical principles.
As it is part of the strategy of these organizations as well as the
German AI strategy (Die Bundesregierung (2018)) to fuel public awareness
and discussion about AI across all parts of society, it could be
beneficial to reach out to people who are at the moment unconcerned
about ethical issues. According to our results, generating at least some
interest as well as trust in the capacities of AI could lead to greater
engagement with ethical design challenges.

The largest share of the German population equally values all ethical
principles and, thus, sets very high standards for ethical AI
development. In fact, this leads to the observation that the bar for
approval of AI systems is very high for this group. However, if the
principles are complied with, ethical AI can lead to high acceptance of
AI. Common characteristics of this group are a high level of education,
young age, and high interest in AI as well as high acceptance of AI.
This group is especially demanding in terms of AI design. This may lead
to a serious problem for AI design. As outlined, some trade-off
decisions must be made eventually, as the simultaneous maximizing of all
ethical principles is very challenging. However, our results suggest
that it will not be easy to satisfy the demands of ethically concerned
people. If some ethical trade-offs are taken, it may very well lead to
reservations against AI.

However, considering only the public perspective in AI development and
implementation might also have serious ramifications. Srivastava et al.
(2019) show regarding algorithmic fairness that the broad public prefers
simple and easy to comprehend algorithms to more complex ones, even if
the complex ones achieved higher factual fairness scores. As AI
technology is complex in its nature, it is possible that many people
will not understand some design settings. In the end, this might lead to
a public demand for systems that are easier to understand. However, it
might very well be that a more thorough design of those systems would
follow ethical principles to an even higher extent. Thus, we highlight
that the public perspective on AI development definitely needs more
attention in science as well as in technology development and
implementation. We emphasize that the public perspective should rather
complement and not dominate other perspectives on AI development and
implementation.

\hypertarget{limitations}{%
\subsection{Limitations}\label{limitations}}

This study has some limitations that need to be recognized. We used an
algorithmic tax fraud identification system as a use case in our study.
Hence, our results are only valid for the specific context. However, as
we wanted to describe preference profiles and cluster characteristics,
we decided to present only one use case. This approach is similar to
studies in the field of fairness perceptions, in which many studies only
use one use case (Grgić-Hlača et al., 2018; Shin, 2021; Shin and Park,
2019). However, public perceptions of AI are highly context dependent.
It might be that importance weights and cluster profiles differ
concerning the particular use case. Therefore, further studies should
test for various use cases simultaneously and compare the results
regarding those contexts. Context-comparing studies have already been
performed for public perception of trust in AI (Araujo et al., 2020) and
threat perceptions (Kieslich et al., 2021).

A general limitation of vignette studies lies in the fact that we used
an artificial use case and that the rating of the system has limited
real-world implications. Without a doubt, more research is needed that
focuses on stakeholders' perspectives on AI systems that are or will be
implemented soon.

Additionally, the survey was conducted only in Germany, and the findings
are thus only valid for the German population. We encourage further
studies that replicate and enhance our study in other countries.
Cross-national studies could detect specific nation patterns regarding
the importance weights and preference profiles of ethical principles.
The comparison to the US, Chinese and UK population would be especially
interesting, since those countries follow a different national strategy
for the development of AI.

\hypertarget{conclusion}{%
\section{Conclusion}\label{conclusion}}

Ethical AI is a major societal challenge. We showed that compliance with
ethical requirements matters for most German citizens. To gain wide
acceptance of AI, these ethical principles have to be taken seriously.
However, we also showed that a notable portion of the German population
does not demand ethical AI implementation. This is critical, as
compliance with ethical AI design is, at least to some level, dependent
on the broad public. If ethical requirements are not explicitly
demanded, one might fear that implementation of those principles might
not be on the highest standard, especially because ethical AI
development is expensive. However, we showed that people who demand high
quality standards are interested in AI as well as aware of the risks.
Thus, to raise demands for ethical AI, it would be a promising way to
raise public interest in the technology.

\hypertarget{acknowledgement}{%
\section*{Acknowledgement}\label{acknowledgement}}
\addcontentsline{toc}{section}{Acknowledgement}

The authors would like to thank Pero Došenović, Marco Lünich and Frank
Marcinkowski for their feedback on the paper. Furthermore, the authors
would like to thank Pascal Kieslich and Max Reichert for their feedback
on the methodological implementation and evaluation of the results.

\hypertarget{funding}{%
\section*{Funding}\label{funding}}
\addcontentsline{toc}{section}{Funding}

This study was conducted as part of the project \emph{Meinungsmonitor
Künstliche Intelligenz} (opinion monitor artificial intelligence). The
project is funded by the \emph{Ministerium für Kultur und Wissenschaft
des Landes Nordrhein-Westfalen} (Ministry of Culture and Science of the
State of North Rhine-Westphalia), Germany.

\hypertarget{references}{%
\section*{References}\label{references}}
\addcontentsline{toc}{section}{References}

\hypertarget{refs}{}
\begin{CSLReferences}{1}{0}
\leavevmode\hypertarget{ref-Acikgoz.2020}{}%
Acikgoz Y, Davison KH, Compagnone M, et al. (2020) Justice perceptions
of artificial intelligence in selection. \emph{International Journal of
Selection and Assessment} 28(4): 399--416. DOI:
\href{https://doi.org/10.1111/ijsa.12306}{10.1111/ijsa.12306}.

\leavevmode\hypertarget{ref-AlgorithmWatch.2020}{}%
Algorithm Watch (ed.) (2020) Automating society report 2020. Available
at: \url{https://automatingsociety.algorithmwatch.org/}.

\leavevmode\hypertarget{ref-Allhutter.2020}{}%
Allhutter D, Mager A, Cech F, et al. (2020) Der AMS-Algorithmus: Eine
soziotechnische Analyse des Arbeitsmarktchancen-Assistenz-System (AMAS). [The AMS Algorithm: A Sociotechnical Analysis of the Labor Market Opportunity Assistance System (AMAS)].
Institut für Technikfolgen-Abschätzung der Österreichischen Akademie der
Wissenschaften (ed.). Wien. Available at:
\url{https://epub.oeaw.ac.at/ita/ita-projektberichte/2020-02.pdf}.

\leavevmode\hypertarget{ref-Ananny.2018}{}%
Ananny M and Crawford K (2018) Seeing without knowing: Limitations of
the transparency ideal and its application to algorithmic
accountability. \emph{New Media {\&} Society} 20(3): 973--989. DOI:
\href{https://doi.org/10.1177/1461444816676645}{10.1177/1461444816676645}.

\leavevmode\hypertarget{ref-Angwin.2016}{}%
Angwin J, Larson J, Mattu S, et al. (2016) Machine bias: There's
software used across the country to predict future criminals. And it's
biased against blacks. May 2016. Available at:
\url{https://www.propublica.org/article/machine-bias-risk-assessments-in-criminal-sentencing}.

\leavevmode\hypertarget{ref-Araujo.2020}{}%
Araujo T, Helberger N, Kruikemeier S, et al. (2020) In AI we trust?
Perceptions about automated decision-making by artificial intelligence.
\emph{AI {\&} Society} 35(6): 611--623. DOI:
\href{https://doi.org/10.1007/s00146-019-00931-w}{10.1007/s00146-019-00931-w}.

\leavevmode\hypertarget{ref-Backhaus.2016}{}%
Backhaus K, Erichson B, Weiber R, et al. (2016) Conjoint-Analyse [Conjoint Analysis]. In:
Backhaus K, Erichson B, Plinke W, et al. (eds) \emph{Multivariate
Analysemethoden}. Berlin, Heidelberg: {Springer Berlin Heidelberg}, pp.
517--567. DOI:
\href{https://doi.org/10.1007/978-3-662-46076-410}{10.1007/978-3-662-46076-410}.

\leavevmode\hypertarget{ref-Beil.2019}{}%
Beil M, Proft I, van Heerden D, et al. (2019) Ethical considerations
about artificial intelligence for prognostication in intensive care.
\emph{Intensive care medicine experimental} 7(70): 1--13. DOI:
\href{https://doi.org/10.1186/s40635-019-0286-6}{10.1186/s40635-019-0286-6}.

\leavevmode\hypertarget{ref-Bennett.2018}{}%
Bennett CJ (2018) The european general data protection regulation: An
instrument for the globalization of privacy standards? \emph{Information
Polity} 23(2): 239--246. DOI:
\href{https://doi.org/10.3233/IP-180002}{10.3233/IP-180002}.

\leavevmode\hypertarget{ref-Berendt.2019}{}%
Berendt B (2019) AI for the common good?! Pitfalls, challenges, and
ethics pen-testing. \emph{Paladyn, Journal of Behavioral Robotics}
10(1): 44--65. DOI:
\href{https://doi.org/10.1515/pjbr-2019-0004}{10.1515/pjbr-2019-0004}.

\leavevmode\hypertarget{ref-Binns.2019}{}%
Binns R and Gallo V (2019) Trade-offs. Available at:
\url{https://ico.org.uk/about-the-ico/news-and-events/ai-blog-trade-offs/}.

\leavevmode\hypertarget{ref-Boehmke.2020}{}%
Boehmke BC and Greenwell B (2020) \emph{Hands-on Machine Learning with
R}. Chapman {\&} hall/CRC the r series. Boca Raton: {CRC Press Taylor
{\&} Francis Group}.

\leavevmode\hypertarget{ref-boyd.2012}{}%
boyd d and Crawford K (2012) Critical questions for big data:
Provocations for a cultural, technological, and scholarly phenomenon.
\emph{Information, Communication {\&} Society} 15(5): 662--679. DOI:
\href{https://doi.org/10.1080/1369118X.2012.678878}{10.1080/1369118X.2012.678878}.

\leavevmode\hypertarget{ref-Burbach.2018}{}%
Burbach L, Nakayama J, Plettenberg N, et al. (2018) User preferences in
recommendation algorithms. In: \emph{Proceedings of the 12th ACM
conference on recommender systems} (eds S Pera, M Ekstrand, X Amatriain,
et al.), New York, NY, USA, 2018, pp. 306--310. ACM. DOI:
\href{https://doi.org/10.1145/3240323.3240393}{10.1145/3240323.3240393}.

\leavevmode\hypertarget{ref-Busuioc.2020}{}%
Busuioc M (2020) Accountable artificial intelligence: Holding algorithms
to account. \emph{Public Administration Review} 29(1): 4349. DOI:
\href{https://doi.org/10.1111/puar.13293}{10.1111/puar.13293}.

\leavevmode\hypertarget{ref-Cath.2018}{}%
Cath C, Wachter S, Mittelstadt B, et al. (2018) Artificial intelligence
and the 'good society': The US, EU, and UK approach. \emph{Science and
engineering ethics} 24(2): 505--528. DOI:
\href{https://doi.org/10.1007/s11948-017-9901-7}{10.1007/s11948-017-9901-7}.

\leavevmode\hypertarget{ref-Diakopoulos.2016}{}%
Diakopoulos N (2016) Accountability in algorithmic decision making.
\emph{Communications of the ACM} 59(2): 56--62. DOI:
\href{https://doi.org/10.1145/2844110}{10.1145/2844110}.

\leavevmode\hypertarget{ref-DieBundesregierung.2018}{}%
Die Bundesregierung (2018) Strategie K{ü}nstliche Intelligenz der
Bundesregierung. [Artificial Intelligence Strategy of the Federal Government]. Available at:
\url{https://www.bundesregierung.de/resource/blob/975226/1550276/3f7d3c41c6e05695741273e78b8039f2/2018-11-15-ki-strategie-data.pdf?download=1}.

\leavevmode\hypertarget{ref-Dietvorst.2015}{}%
Dietvorst BJ, Simmons JP and Massey C (2015) Algorithm aversion: People
erroneously avoid algorithms after seeing them err. \emph{Journal of
Experimental Psychology} 144(1): 114--126. DOI:
\href{https://doi.org/10.1037/xge0000033}{10.1037/xge0000033}.

\leavevmode\hypertarget{ref-Elzayn.2020}{}%
Elzayn H and Fish B (2020) The effects of competition and regulation on
error inequality in data-driven markets. In: \emph{Proceedings of the
2020 conference on fairness, accountability, and transparency} (eds M
Hildebrandt, C Castillo, E Celis, et al.), New York, NY, USA, 2020, pp.
669--679. {NY: ACM}. DOI:
\href{https://doi.org/10.1145/3351095.3372842}{10.1145/3351095.3372842}.

\leavevmode\hypertarget{ref-Eslami.2018}{}%
Eslami M, Krishna Kumaran SR, Sandvig C, et al. (2018) Communicating
algorithmic process in online behavioral advertising. In:
\emph{Proceedings of the 2018 CHI conference on human factors in
computing systems} (eds R Mandryk, M Hancock, M Perry, et al.), New
York, NY, USA, 2018, pp. 1--13. ACM. DOI:
\href{https://doi.org/10.1145/3173574.3174006}{10.1145/3173574.3174006}.

\leavevmode\hypertarget{ref-EuropeanCommission.2019}{}%
European Commission (2019) Ethics guidelines for trustworthy AI.
Available at:
\url{https://digital-strategy.ec.europa.eu/en/library/ethics-guidelines-trustworthy-ai}.

\leavevmode\hypertarget{ref-EuropeanCommission.2020}{}%
European Commission (2020) White paper on artificial intelligence - a
European approach to excellence and trust. Available at:
\url{https://ec.europa.eu/info/sites/info/files/commission-white-paper-artificial-intelligence-feb2020_en.pdf}.

\leavevmode\hypertarget{ref-EuropeanCommission.2021}{}%
European Commission (2021) Proposal for a regulation of the European
Parliament and of the council: Laying down harmonised rules on
artificial intelligence (artificial intelligence act) and amending
certain union legislative acts. Available at:
\url{https://ec.europa.eu/newsroom/dae/document.cfm?doc_id=75788}.

\leavevmode\hypertarget{ref-Field.2011}{}%
Field A (2011) \emph{Discovering Statistics Using SPSS: (And Sex and
Drugs and Rock 'n' Roll)}. 3. ed., reprinted. Los Angeles, Calif.: Sage.

\leavevmode\hypertarget{ref-Floridi.2018}{}%
Floridi L, Cowls J, Beltrametti M, et al. (2018) AI4People - an ethical
framework for a good AI society: Opportunities, risks, principles, and
recommendations. \emph{Minds and machines} 28(4): 689--707. DOI:
\href{https://doi.org/10.1007/s11023-018-9482-5}{10.1007/s11023-018-9482-5}.

\leavevmode\hypertarget{ref-Graham.2019}{}%
Graham S, Depp C, Lee EE, et al. (2019) Artificial intelligence for
mental health and mental illnesses: An overview. \emph{Current
psychiatry reports} 21(11): 1--18. DOI:
\href{https://doi.org/10.1007/s11920-019-1094-0}{10.1007/s11920-019-1094-0}.

\leavevmode\hypertarget{ref-Green.2001}{}%
Green PE, Krieger AM and Wind Y (2001) Thirty years of conjoint
analysis: Reflections and prospects. \emph{Interfaces}
31(3{\_}supplement): S56--S73. DOI:
\href{https://doi.org/10.1287/inte.31.3s.56.9676}{10.1287/inte.31.3s.56.9676}.

\leavevmode\hypertarget{ref-GrgicHlaca.2018}{}%
Grgić-Hlača N, Redmiles EM, Gummadi KP, et al. (2018) Human perceptions
of fairness in algorithmic decision making. In: \emph{Proceedings of the
2018 world wide web conference on world wide web - WWW '18} (eds P-A
Champin, F Gandon, M Lalmas, et al.), New York, New York, USA, 2018, pp.
903--912. {ACM Press}. DOI:
\href{https://doi.org/10.1145/3178876.3186138}{10.1145/3178876.3186138}.

\leavevmode\hypertarget{ref-GrgicHlaca.2020}{}%
Grgić-Hlača N, Weller A and Redmiles EM (2020) Dimensions of diversity
in human perceptions of algorithmic fairness. Available at:
\url{http://arxiv.org/pdf/2005.00808v1}.

\leavevmode\hypertarget{ref-Hagendorff.2020}{}%
Hagendorff T (2020) The ethics of AI ethics: An evaluation of
guidelines. \emph{Minds and Machines} 30(1): 99--120. DOI:
\href{https://doi.org/10.1007/s11023-020-09517-8}{10.1007/s11023-020-09517-8}.

\leavevmode\hypertarget{ref-Hancock.2011}{}%
Hancock PA, Billings DR, Schaefer KE, et al. (2011) A meta-analysis of
factors affecting trust in human-robot interaction. \emph{Human factors}
53(5): 517--527. DOI:
\href{https://doi.org/10.1177/0018720811417254}{10.1177/0018720811417254}.

\leavevmode\hypertarget{ref-Hartigan.1979}{}%
Hartigan JA and Wong MA (1979) Algorithm AS 136: A k-means clustering
algorithm. \emph{Applied Statistics} 28(1): 100--108. DOI:
\href{https://doi.org/10.2307/2346830}{10.2307/2346830}.

\leavevmode\hypertarget{ref-Hardle.2015}{}%
Härdle WK and Simar L (2015) Conjoint measurement analysis. In: Härdle
WK and Simar L (eds) \emph{Applied Multivariate Statistical Analysis}.
Berlin, Heidelberg: {Springer Berlin Heidelberg}, pp. 473--486. DOI:
\href{https://doi.org/10.1007/978-3-662-45171-718}{10.1007/978-3-662-45171-718}.

\leavevmode\hypertarget{ref-Helberger.2020}{}%
Helberger N, Araujo T and Vreese CH de (2020) Who is the fairest of them
all? Public attitudes and expectations regarding automated
decision-making. \emph{Computer Law {\&} Security Review} 39: 105456.
DOI:
\href{https://doi.org/10.1016/j.clsr.2020.105456}{10.1016/j.clsr.2020.105456}.

\leavevmode\hypertarget{ref-InstitutfurdenoffentlichenSektor.2019}{}%
Institut für den öffentlichen Sektor (ed.) (2019) Land Hessen:
K{ü}nstliche intelligenz in der Steuerfahndung. [State of Hesse: Artificial intelligence in tax investigation]. Available at:
\url{https://publicgovernance.de/html/de/8497.htm}.

\leavevmode\hypertarget{ref-Jhaver.2018}{}%
Jhaver S, Karpfen Y and Antin J (2018) Algorithmic anxiety and coping
strategies of airbnb hosts. In: \emph{Proceedings of the 2018 CHI
conference on human factors in computing systems} (eds R Mandryk, M
Hancock, M Perry, et al.), New York, NY, USA, 2018, pp. 1--12. ACM. DOI:
\href{https://doi.org/10.1145/3173574.3173995}{10.1145/3173574.3173995}.

\leavevmode\hypertarget{ref-Jobin.2019}{}%
Jobin A, Ienca M and Vayena E (2019) The global landscape of AI ethics
guidelines. \emph{Nature Machine Intelligence} 1(9): 389--399. DOI:
\href{https://doi.org/10.1038/s42256-019-0088-2}{10.1038/s42256-019-0088-2}.

\leavevmode\hypertarget{ref-Kelly.2021}{}%
Kelly A (2021) A tale of two algorithms: The appeal and repeal of
calculated grades systems in england and ireland in 2020. \emph{British
Educational Research Journal}. DOI:
\href{https://doi.org/10.1002/berj.3705}{10.1002/berj.3705}.

\leavevmode\hypertarget{ref-Kieslich.2020}{}%
Kieslich K, Starke C, Došenović P, et al. (2020) Artificial intelligence
and discrimination. Available at:
\url{https://www.cais.nrw/en/factsheet-2-ai-discrimination/}.

\leavevmode\hypertarget{ref-Kieslich.2021}{}%
Kieslich K, Lünich M and Marcinkowski F (2021) The threats of artificial
intelligence scale (TAI). \emph{International Journal of Social
Robotics}: 216. DOI:
\href{https://doi.org/10.1007/s12369-020-00734-w}{10.1007/s12369-020-00734-w}.

\leavevmode\hypertarget{ref-Kizilcec.2016}{}%
Kizilcec RF (2016) How much information? In: \emph{Proceedings of the
2016 CHI conference on human factors in computing systems} (eds J Kaye,
A Druin, C Lampe, et al.), New York, NY, USA, 2016, pp. 2390--2395. ACM.
DOI:
\href{https://doi.org/10.1145/2858036.2858402}{10.1145/2858036.2858402}.

\leavevmode\hypertarget{ref-Kobis.2021}{}%
Köbis N, Starke C and Rahwan I (2021) Artificial intelligence as an
anti-corruption tool (AI-ACT): Potentials and pitfalls for top-down and
bottom-up approaches. Available at:
\url{https://arxiv.org/abs/2102.11567}.

\leavevmode\hypertarget{ref-Kochling.2020}{}%
Köchling A and Wehner MC (2020) Discriminated by an algorithm: A
systematic review of discrimination and fairness by algorithmic
decision-making in the context of HR recruitment and HR development.
\emph{Business Research} 13(3): 795--848. DOI:
\href{https://doi.org/10.1007/s40685-020-00134-w}{10.1007/s40685-020-00134-w}.

\leavevmode\hypertarget{ref-Lau.2018}{}%
Lau J, Zimmerman B and Schaub F (2018) Alexa, are you listening?
\emph{Proceedings of the ACM on Human-Computer Interaction} 2(CSCW):
1--31. DOI: \href{https://doi.org/10.1145/3274371}{10.1145/3274371}.

\leavevmode\hypertarget{ref-Lee.2018}{}%
Lee MK (2018) Understanding perception of algorithmic decisions:
Fairness, trust, and emotion in response to algorithmic management.
\emph{Big Data {\&} Society} 5(1): 1--16. DOI:
\href{https://doi.org/10.1177/2053951718756684}{10.1177/2053951718756684}.

\leavevmode\hypertarget{ref-Lee.2017}{}%
Lee MK, Kim JT and Lizarondo L (2017) A human-centered approach to
algorithmic services: Considerations for fair and motivating smart
community service management that allocates donations to non-profit
organizations. In: \emph{Proceedings of the 2017 CHI conference on human
factors in computing systems} (eds G Mark, S Fussell, C Lampe, et al.),
New York, NY, USA, 2017, pp. 3365--3376. ACM. DOI:
\href{https://doi.org/10.1145/3025453.3025884}{10.1145/3025453.3025884}.

\leavevmode\hypertarget{ref-Leventhal.1980}{}%
Leventhal GS (1980) What should be done with equity theory? In: Gergen
KJ, Greenberg MS, and Willis RH (eds) \emph{Social Exchange}. Boston,
MA: {Springer US}, pp. 27--55. DOI:
\href{https://doi.org/10.1007/978-1-4613-3087-52}{10.1007/978-1-4613-3087-52}.

\leavevmode\hypertarget{ref-Liang.2017}{}%
Liang Y and Lee SA (2017) Fear of autonomous robots and artificial
intelligence: Evidence from national representative data with
probability sampling. \emph{International Journal of Social Robotics}
9(3): 379--384. DOI:
\href{https://doi.org/10.1007/s12369-017-0401-3}{10.1007/s12369-017-0401-3}.

\leavevmode\hypertarget{ref-Lim.2011}{}%
Lim BY and Dey AK (2011) Investigating intelligibility for uncertain
context-aware applications. In: \emph{Proceedings of the 13th
international conference on ubiquitous computing - UbiComp '11} (eds J
Landay, Y Shi, DJ Patterson, et al.), New York, New York, USA, 2011, p.
415. {ACM Press}. DOI:
\href{https://doi.org/10.1145/2030112.2030168}{10.1145/2030112.2030168}.

\leavevmode\hypertarget{ref-Liu.2009}{}%
Liu H and Priest S (2009) Understanding public support for stem cell
research: Media communication, interpersonal communication and trust in
key actors. \emph{Public Understanding of Science} 18(6): 704--718. DOI:
\href{https://doi.org/10.1177/0963662508097625}{10.1177/0963662508097625}.

\leavevmode\hypertarget{ref-Machanavajjhala.2011}{}%
Machanavajjhala A, Korolova A and Sarma AD (2011) Personalized social
recommendations - accurate or private? \emph{Proceedings of the VLDB
Endowment (PVLDB)} 4(7). Available at:
\url{http://arxiv.org/pdf/1105.4254v1}.

\leavevmode\hypertarget{ref-Marcinkowski.2020}{}%
Marcinkowski F, Kieslich K, Starke C, et al. (2020) Implications of AI
(un-)fairness in higher education admissions. In: \emph{Proceedings of
the 2020 conference on fairness, accountability, and transparency} (eds
M Hildebrandt, C Castillo, E Celis, et al.), New York, NY, USA, 2020,
pp. 122--130. {NY: ACM}. DOI:
\href{https://doi.org/10.1145/3351095.3372867}{10.1145/3351095.3372867}.

\leavevmode\hypertarget{ref-McNamara.2018}{}%
McNamara A, Smith J and Murphy-Hill E (2018) Does ACM's code of ethics
change ethical decision making in software development? In:
\emph{Proceedings of the 2018 26th ACM joint meeting on european
software engineering conference and symposium on the foundations of
software engineering} (eds GT Leavens, A Garcia, and CS Păsăreanu), New
York, NY, USA, 2018, pp. 729--733. ACM. DOI:
\href{https://doi.org/10.1145/3236024.3264833}{10.1145/3236024.3264833}.

\leavevmode\hypertarget{ref-MeinungsmonitorKunstlicheIntelligenz.2021}{}%
Meinungsmonitor Künstliche Intelligenz (2021) Ein Instrument zur
kontinuierlichen Beobachtung {ö}ffentlicher und ver{ö}ffentlichter
Meinung zu KI. [An instrument for the continuous monitoring of public and published opinion on AI]. Available at: \url{https://www.cais.nrw/memoki/}.

\leavevmode\hypertarget{ref-Miller.2018}{}%
Miller AP (2018) Want less-biased decisions? Use algorithms.
\emph{Harvard Business Review}. Available at:
\url{https://hbr.org/2018/07/want-less-biased-decisions-use-algorithms}.

\leavevmode\hypertarget{ref-Nagtegaal.2021}{}%
Nagtegaal R (2021) The impact of using algorithms for managerial
decisions on public employees' procedural justice. \emph{Government
Information Quarterly} 38(1): 101536. DOI:
\href{https://doi.org/10.1016/j.giq.2020.101536}{10.1016/j.giq.2020.101536}.

\leavevmode\hypertarget{ref-Neuhaus.2019}{}%
Neuhaus R, Laschke M, Theofanou-Fülbier D, et al. (2019) Exploring the
impact of transparency on the interaction with an in-car digital AI
assistant. In: \emph{Proceedings of the 11th international conference on
automotive user interfaces and interactive vehicular applications:
Adjunct proceedings} (eds CP Janssen, SF Donker, LL Chuang, et al.), New
York, NY, USA, 2019, pp. 450--455. ACM. DOI:
\href{https://doi.org/10.1145/3349263.3351325}{10.1145/3349263.3351325}.

\leavevmode\hypertarget{ref-OECD.2021}{}%
OECD (2021) Recommendation of the council on artificial intelligence.
Available at:
\url{https://legalinstruments.oecd.org/en/instruments/OECD-LEGAL-0449}.

\leavevmode\hypertarget{ref-PewResearchCenter.2018}{}%
Pew Research Center (2018) Public attitudes towards computer algorithms.
Available at:
\url{https://www.pewresearch.org/internet/2018/11/16/public-attitudes-toward-computer-algorithms/}.

\leavevmode\hypertarget{ref-Robinette.2016}{}%
Robinette P, Li W, Allen R, et al. (2016) Overtrust of robots in
emergency evacuation scenarios. In: \emph{2016 11th ACM/IEEE
international conference on human-robot interaction (HRI)} (eds C
Bartneck, Y Nagai, A Paiva, et al.), New York, NY, USA, 2016, pp.
101--108. {NY: ACM}. DOI:
\href{https://doi.org/10.1109/HRI.2016.7451740}{10.1109/HRI.2016.7451740}.

\leavevmode\hypertarget{ref-Robinette.2017}{}%
Robinette P, Howard AM and Wagner AR (2017) Effect of robot performance
on human--robot trust in time-critical situations. \emph{IEEE
Transactions on Human-Machine Systems} 47(4): 425--436. DOI:
\href{https://doi.org/10.1109/THMS.2017.2648849}{10.1109/THMS.2017.2648849}.

\leavevmode\hypertarget{ref-Saha.2020}{}%
Saha D, Schumann C, McElfresh DC, et al. (2020) Human comprehension of
fairness in machine learning. In: \emph{Proceedings of the AAAI/ACM
conference on AI, ethics, and society} (eds A Markham, J Powles, T
Walsh, et al.), New York, NY, USA, 2020, p. 152. ACM. DOI:
\href{https://doi.org/10.1145/3375627.3375819}{10.1145/3375627.3375819}.

\leavevmode\hypertarget{ref-Salem.2015}{}%
Salem M, Lakatos G, Amirabdollahian F, et al. (2015) Would you trust a
(faulty) robot? In: \emph{Proceedings of the tenth annual ACM/IEEE
international conference on human-robot interaction} (eds JA Adams, W
Smart, B Mutlu, et al.), New York, NY, USA, 2015, pp. 141--148. ACM.
DOI:
\href{https://doi.org/10.1145/2696454.2696497}{10.1145/2696454.2696497}.

\leavevmode\hypertarget{ref-Shin.2020}{}%
Shin D (2020) User perceptions of algorithmic decisions in the
personalized AI system: Perceptual evaluation of fairness,
accountability, transparency, and explainability. \emph{Journal of
Broadcasting {\&} Electronic Media} 64(4): 541--565. DOI:
\href{https://doi.org/10.1080/08838151.2020.1843357}{10.1080/08838151.2020.1843357}.

\leavevmode\hypertarget{ref-Shin.2021}{}%
Shin D (2021) The effects of explainability and causability on
perception, trust, and acceptance: Implications for explainable AI.
\emph{International Journal of Human-Computer Studies} 146: 102551. DOI:
\href{https://doi.org/10.1016/j.ijhcs.2020.102551}{10.1016/j.ijhcs.2020.102551}.

\leavevmode\hypertarget{ref-Shin.2019}{}%
Shin D and Park YJ (2019) Role of fairness, accountability, and
transparency in algorithmic affordance. \emph{Computers in Human
Behavior} 98: 277--284. DOI:
\href{https://doi.org/10.1016/j.chb.2019.04.019}{10.1016/j.chb.2019.04.019}.

\leavevmode\hypertarget{ref-Shin.2020b}{}%
Shin D, Zhong B and Biocca FA (2020) Beyond user experience: What
constitutes algorithmic experiences? \emph{International Journal of
Information Management} 52(3): 102061. DOI:
\href{https://doi.org/10.1016/j.ijinfomgt.2019.102061}{10.1016/j.ijinfomgt.2019.102061}.

\leavevmode\hypertarget{ref-Shneiderman.2020}{}%
Shneiderman B (2020) Human-centered artificial intelligence: Reliable,
safe {\&} trustworthy. \emph{International Journal of Human--Computer
Interaction} 36(6): 495--504. DOI:
\href{https://doi.org/10.1080/10447318.2020.1741118}{10.1080/10447318.2020.1741118}.

\leavevmode\hypertarget{ref-Srivastava.2019}{}%
Srivastava M, Heidari H and Krause A (2019) Mathematical notions vs.
Human perception of fairness. In: \emph{Proceedings of the 25th ACM
SIGKDD international conference on knowledge discovery {\&} data mining}
(eds A Teredesai, V Kumar, Y Li, et al.), New York, NY, USA, 2019, pp.
2459--2468. {NY: ACM}. DOI:
\href{https://doi.org/10.1145/3292500.3330664}{10.1145/3292500.3330664}.

\leavevmode\hypertarget{ref-Starke.2020}{}%
Starke C and Lünich M (2020) Artificial intelligence for political
decision-making in the european union: Effects on citizens' perceptions
of input, throughput, and output legitimacy. \emph{Data {\&} Policy}
2(e16). DOI:
\href{https://doi.org/10.1017/dap.2020.19}{10.1017/dap.2020.19}.

\leavevmode\hypertarget{ref-Starke.2021}{}%
Starke C, Baleis J, Keller B, et al. (2021) Fairness perceptions of
algorithmic decision-making: A systematic review of the empirical
literature. Available at: \url{http://arxiv.org/pdf/2103.12016v1}.

\leavevmode\hypertarget{ref-Vallejos.2017}{}%
Vallejos EP, Koene A, Portillo V, et al. (2017) Young people's policy
recommendations on algorithm fairness. In: \emph{Proceedings of the 2017
ACM on web science conference} (eds P Fox, D McGuinness, L Poirer, et
al.), New York, NY, USA, 2017, pp. 247--251. {NY: ACM}. DOI:
\href{https://doi.org/10.1145/3091478.3091512}{10.1145/3091478.3091512}.

\leavevmode\hypertarget{ref-Vitale.2018}{}%
Vitale J, Tonkin M, Herse S, et al. (2018) Be more transparent and users
will like you. In: \emph{Proceedings of the 2018 ACM/IEEE international
conference on human-robot interaction} (eds T Kanda, S Ŝabanović, G
Hoffman, et al.), New York, NY, USA, 2018, pp. 379--387. {NY: ACM}. DOI:
\href{https://doi.org/10.1145/3171221.3171269}{10.1145/3171221.3171269}.

\leavevmode\hypertarget{ref-Wang.2020}{}%
Wang R, Harper FM and Zhu H (2020) Factors influencing perceived
fairness in algorithmic decision-making. In: \emph{Proceedings of the
2020 CHI conference on human factors in computing systems} (eds R
Bernhaupt, F Mueller, D Verweij, et al.), New York, NY, USA, 2020, pp.
1--14. ACM. DOI:
\href{https://doi.org/10.1145/3313831.3376813}{10.1145/3313831.3376813}.

\leavevmode\hypertarget{ref-Webb.2018}{}%
Webb H, Koene A, Patel M, et al. (2018) Multi-stakeholder dialogue for
policy recommendations on algorithmic fairness. In: \emph{Proceedings of
the 9th international conference on social media and society} (eds A
Gruzd, J Jacobsen, P Mai, et al.), NY, 2018, pp. 395--399. {NY: ACM}.
DOI:
\href{https://doi.org/10.1145/3217804.3217952}{10.1145/3217804.3217952}.

\leavevmode\hypertarget{ref-Yeasmin.2019}{}%
Yeasmin S (2019) Benefits of artificial intelligence in medicine. In:
\emph{2nd international conference on computer applications {\&}
information security (ICCAIS)}, 2019, pp. 1--6. IEEE. DOI:
\href{https://doi.org/10.1109/CAIS.2019.8769557}{10.1109/CAIS.2019.8769557}.

\leavevmode\hypertarget{ref-Zheng.2018}{}%
Zheng S, Apthorpe N, Chetty M, et al. (2018) User perceptions of smart
home IoT privacy. \emph{Proceedings of the ACM on Human-Computer
Interaction} 2(CSCW): 1--20. DOI:
\href{https://doi.org/10.1145/3274469}{10.1145/3274469}.

\leavevmode\hypertarget{ref-Zou.2018}{}%
Zou J and Schiebinger L (2018) AI can be sexist and racist - it's time
to make it fair. \emph{Nature} 559(7714): 324--326. DOI:
\href{https://doi.org/10.1038/d41586-018-05707-8}{10.1038/d41586-018-05707-8}.

\end{CSLReferences}

\hypertarget{appendix}{%
\section*{Appendix}\label{appendix}}
\addcontentsline{toc}{section}{Appendix}

Table 4 depicts the orthoplan for the conjoint analysis. Table 5
displays the item wording.

\begin{table}[H]

\caption{\label{tab:unnamed-chunk-7}Orthoplan}
\centering
\resizebox{\linewidth}{!}{
\begin{tabular}[t]{llllllll}
\toprule
Karten-ID & Explainability & Fairness & Security & Accountability & Accuracy & Privacy & Control\\
\midrule
A & Yes & Yes & Yes & No & No & No & No\\
B & Yes & No & No & Yes & No & No & Yes\\
C & No & No & Yes & No & Yes & No & Yes\\
D & Yes & Yes & Yes & Yes & Yes & Yes & Yes\\
E & No & Yes & No & No & No & Yes & Yes\\
\addlinespace
F & Yes & No & No & No & Yes & Yes & No\\
G & No & No & Yes & Yes & No & Yes & No\\
H & No & Yes & No & Yes & Yes & No & No\\
\bottomrule
\end{tabular}}
\end{table}

\newpage

\begin{table}[H]

\caption{\label{tab:unnamed-chunk-8}Item Wording}
\centering
\begin{tabu} to \linewidth {>{\raggedright}X}
\toprule
Item Wording\\
\midrule
\addlinespace[0.3em]
\multicolumn{1}{l}{\textbf{Interest in AI}}\\
\em{\hspace{1em}How much do the following statements apply to you?}\\
\hspace{1em}I follow processes around artificial intelligence with great curiosity.\\
\hspace{1em}In general, I am very interested in artificial intelligence.\\
\hspace{1em}I read articles about artificial intelligence with great attention.\\
\hspace{1em}I watch or listen to articles about Artificial Intelligence with great interest.\\
\addlinespace[0.3em]
\multicolumn{1}{l}{\textbf{Acceptance of AI}}\\
\em{\hspace{1em}Are you more in favor of or against the use of artificial intelligence...}\\
\hspace{1em}...at banks and savings banks?\\
\hspace{1em}...in the health care sector?\\
\hspace{1em}...in industrial production?\\
\hspace{1em}...in traffic?\\
\hspace{1em}...in personal everyday life?\\
\hspace{1em}...in schools and universities?\\
\hspace{1em}...in public administration?\\
\hspace{1em}...in political decisions?\\
\hspace{1em}...in court?\\
\hspace{1em}...in police and security agencies?\\
\hspace{1em}...in journalism?\\
\hspace{1em}...in personnel decisions?\\
\hspace{1em}...in insurance companies?\\
\hspace{1em}...in the military?\\
\addlinespace[0.3em]
\multicolumn{1}{l}{\textbf{Risk awareness of AI}}\\
\em{\hspace{1em}How big do you think is the risk posed by artificial \vphantom{1}intelligence...}\\
\hspace{1em}...for our society as a \vphantom{1}whole?\\
\hspace{1em}...for you \vphantom{1}personally?\\
\hspace{1em}...for your family, friends or \vphantom{1}acquaintances?\\
\addlinespace[0.3em]
\multicolumn{1}{l}{\textbf{Opportunity awareness of AI}}\\
\em{\hspace{1em}How big do you think is the risk posed by artificial intelligence...}\\
\hspace{1em}...for our society as a whole?\\
\hspace{1em}...for you personally?\\
\hspace{1em}...for your family, friends or acquaintances?\\
\addlinespace[0.3em]
\multicolumn{1}{l}{\textbf{Trust in AI}}\\
\em{\hspace{1em}How much do you trust artificial intelligence systems already today to...}\\
\hspace{1em}...recognize patterns in large data sets\\
\hspace{1em}...make correct predictions about future developments\\
\hspace{1em}...make appropriate recommendations for human actions\\
\hspace{1em}...make high-quality decisions itself\\
\bottomrule
\end{tabu}
\end{table}

\bibliographystyle{unsrt}
\bibliography{references.bib}

\end{document}